\documentclass[]{aa}
\usepackage{graphicx}
\usepackage{graphics}
\usepackage{epsfig}
\newcommand{\be}{\begin{eqnarray}}
\newcommand{\ee}{\end{eqnarray}}
\begin{document}
\title{Precession of neutrino-cooled accretion disks in gamma-ray burst engines}
\author{Mat\'{\i}as M. Reynoso\inst{1,}\thanks{Fellow of CONICET, Argentina}, Gustavo E. Romero\inst{2,3,}\thanks{Member of CONICET, Argentina}, and Oscar A. Sampayo
\inst{1,}$^{\star\star}$}
\offprints{M.M. Reynoso\\  \email{mreynoso@mdp.edu.ar}}
\titlerunning{Accretion disk precession in GRBs}
\authorrunning{M.M. Reynoso, G.E. Romero \& O.A. Sampayo}
\institute{Departamento de F\'{\i}sica, Facultad de Ciencias Exactas y Naturales, Universidad Nacional de Mar del Plata, Funes 3350, (7600) Mar del Plata, Argentina \and Instituto Argentino de Radioastronom\'{\i}a, C.C.5,
(1894) Villa Elisa, Buenos Aires, Argentina \and Facultad de
Ciencias Astron\'omicas y Geof\'{\i}sicas, Universidad Nacional de La Plata, Paseo del Bosque, (1900) La Plata, Argentina}
\date{Received 22 November 2005 / Accepted 17 March 2006}

   {} 
  \abstract{We study the precession of accretion disks in the context
     of gamma-ray burst inner engines.
   With an accretion disk model that
allows for neutrino cooling, we evaluate the possible periods of
disk precession and nutation due to the Lense-Thirring effect.
   Assuming jet ejection perpendicular to the disk midplane and a
typical intrinsic time dependence for the burst, we find possible
gamma-ray light curves with a temporal microstructure similar to
what is observed in some subsamples.
   We conclude that the precession and
nutation of a neutrino-cooled accretion disk in the burst engine
might be responsible for some events, especially those with a slow
rise and a fast decay.

}

\maketitle

\section{Introduction}
Currently favored models for gamma-ray burst (GRB) central
engines, such as a failed supernova or `collapsar' (Woosley 1993),
the merging of compact objects (Paczynski 1986, Eichler et al.
1989, Narayan et al. 1992), and common-envelope evolution in
compact binaries (Fryer \& Woosley 1998), lead to the formation of
a transient hot and dense accretion disk around
a black hole of a few solar masses. Such a disk can be dominantly
cooled through neutrino losses. The accretion of matter, with a
rate $\sim 0.1-1$ $M_\odot \rm s^{-1}$, would power the burst, and
the energy ejected through relativistic jets is expected to
account for the observed GRB light curves. These curves display a
wide variety of forms with timescales from milliseconds to
minutes. The usual interpretation of this temporal structure is in
terms of shocks that convert bulk kinetic energy into internal
energy of the particles, which then cool through synchrotron and
inverse Compton emission. The shocks can be internal to the jet
and due to colliding shells of different Lorentz factors (e.g.
Kobayashi, Piran \& Sari 1997; Daigne \& Mochkovitch 1998; Guetta,
Spada \& Waxman 2001) or the result of interactions with the
ambient medium (e.g. Heinz \& Begelman 1999; Fenimore, Madras \&
Nayakshin 1996). Among the observed light curves, however, there
are some that are hard
to explain 
using the standard model, like those with a slow rise and a fast
decay (Romero et al. 1999).

In this paper, we study the possible precession of dense accretion
disks, hence of the jets produced in GRBs. We find that for
typical sets of parameters, like a black hole with a mass
$M_\mathrm{bh}=3M_\odot$, a dimensionless spin parameter $a_*= 0.9
$, and accretion rates in the range $\dot{M}=0.1-1 M_\odot \rm
s^{-1}$, spin-induced precession can occur in the neutrino-cooled
disk. The precession of the disk is transmitted to the
relativistic jets, which results in the peculiar temporal
microstructure of some GRB light curves.
In this way we provide a new basis for the precession of GRB jets
(see also Blackman et al. 1996; Portegies et al. 1999; Fargion
1999, 2003) and, in particular, a possible origin of light curves
with slower rises than decays.

The structure of this work is as follows. In Sect. 2, we describe
the basic equations of a neutrino-cooled accretion disk within the
model considered and obtain the disk structure. In Sect. 3, we
analyze the precession of the disk due to the Lense-Thirring
effect and how it relates to the disk density and outer radius. In
Sect. 4 we focus on the precession of disks cooled by neutrino
emission with general properties such as those outlined in Sect.
2.
Finally, we present a discussion with examples in Sect. 5 and a
summary with our final comments in Sect. 6.

\section{Neutrino-cooled accretion disks}

A neutrino-cooled accretion disk occurs when rotating matter
around a compact object (which we take to be a black hole)
achieves the transport of its angular momentum outwards, and as it
falls in the potential well, it liberates most of its
gravitational energy as neutrinos. Central black holes are rapidly
rotating in most candidate GRB engine models, and the structure of
accretion disks around such compact objects has been extensively
studied in different contexts (e.g. Novikov \& Thorne 1973;
Riffert \& Herold 1995; Artemova et al. 1996; Belovorodov 1998;
Setiawan et al. 2004).

However, to take neutrino losses and transfer into account, we
implement a steady-state disk model similar to the one presented
by Di Matteo et al. (2002), but with the necessary relativistic
correction factors according to Riffert \& Herold (1995), due to
the presence of a Kerr black hole.
As mentioned by Di Matteo  et al. (2002) and Popham et al. (1999),
although in GRB's engines the accretion rate may vary, it is
expected to vary significantly only in the outer disk. Hence it is
a good approximation to study the inner neutrino-cooled disk
assuming a constant accretion rate in order to obtain its rough
properties.
We now describe the basic equations on which our accretion model
is based. Conservation of mass implies that the accretion rate is
 \be
  \dot{M}=-2 \pi r v_r \Sigma,\label{masscnsv}
 \ee where $\Sigma=2 \rho H$
is the surface density and $H$ is the disk's half thickness. The
angular momentum of the disk material is diminished in magnitude
due to the action of viscosity. Thus, conservation of angular
momentum in the direction perpendicular to the disk implies
 \be f_\phi(2\pi r
 \cdot 2H )\cdot r= \frac{\Delta L}{\Delta t},\label{angcnsv}
 \ee
where $f_\phi$ is the viscous stress and $L$ the disk angular
momentum component in the direction perpendicular to the disk
midplane. Then, $\Delta L/\Delta t= \dot{M} \Omega r^2$, where
$\Omega = \sqrt{GM_\mathrm{bh}/r^3}$ is the Keplerian angular
velocity. We adopt the usual prescription for viscosity: $f_\phi =
-t_{\phi r}=\alpha P$, where $t_{\phi r}$ is the $\phi-r$ component
of the viscous stress tensor, $\alpha$ the dimensionless viscosity
parameter (Shakura \& Sunyaev 1973), and $P$ the total pressure. The
latter includes the contributions of gas, radiation plus
electron-positron pairs, degeneracy, and neutrino pressure,
respectively:
 \be
 P=\rho \frac{kT}{m_n}\left(\frac{1+3X_{\rm nuc}}{4}\right)+  \frac{11}{12}aT^4+
 K\left(\frac{\rho}{2}\right)^{4/3}
 +\frac{u_\nu}{3},\label{Pest}
 \ee
where
 \be K=\frac{2\pi h c}{3}\left( \frac{3}{8\pi
m_n}\right)^{\frac{4}{3}}, \ee
 $X_{\rm nuc}$ is the mass fraction
of free nucleons specified below, $m_n$ is the nucleon mass, $a$
the radiation density constant and $u_\nu$ is the neutrino energy
density given by
$u_\nu=\sum_i(\tau_{\nu_i}/2+1/\sqrt{3})/(\tau_{\nu_i}/2+1/\sqrt{3}+1/(3\tau_{a,\nu_i}))$
(Phopam \& Narayan 1995), with $\tau_i$ as the neutrino optical
depth for the flavor $i$ (see below).

 The rate at which energy (entropy) per volume unit is generated by
viscosity for a Keplerian disk is $\dot{Q}^+_{\rm vis}=-f_\phi
t_{\phi r }/\nu$, where $\nu$ is the kinematic viscosity coefficient
(e.g. Shaphiro \& Teukolsky 1983), so that the heating rate is found
to be
 \be
 \dot{Q}^+_\mathrm{vis}=\frac{3 \dot{M}G  M_{\rm bh} }{8\pi H r^3}.
 \ee

Solutions for the disk in thermal equilibrium arise when the energy
balance is achieved (e.g., Kohri \& Mineshige 2002):
 \be
 Q_\mathrm{vis}^+= Q^-,\label{bal}
 \ee
where $Q^+_\mathrm{vis}=\dot{Q}^+_\mathrm{vis}H$
is the viscous heating rate integrated over the half thickness,
and the corresponding cooling term consists in turn of four
contributions, \be Q^-= Q_\mathrm{rad}^-+ Q_\mathrm{photo}^-
+Q_\mathrm{adv}^-+Q_\mathrm{\nu}^-.\ee The radiative cooling $
Q_\mathrm{rad}^-$ is negligible as compared with the other sources
of cooling in the range of high temperatures and densities that we
deal with. The photodissociation of heavy nuclei constitutes a
cooling mechanism, with a rate approximately given by
 \be
  Q_\mathrm{photo}^-=10^{29} \rho_{10}v_r
\frac{dX_{\rm nuc}}{dr}H \ \rm{erg \ } \rm{cm}^{-2}{\rm s}^{-1},
 \ee
where $\rho_{10}=\rho/10^{10}{\rm g \ cm}^{-3}$ and \be X_{\rm
nuc} \approx 34.8 \rho_{10}^{-3/4}T_{11}^{9/8}{\rm
exp}(-0.61/T_{11}),\ee with $T_{11}=T/10^{11}{\rm K}$ (e.g. Popham
et al. 1999).

We approximate the advective cooling term as in Di Mateo et al.
(2002; see also Narayan \& Yi 1994)
 \be
 Q_\mathrm{adv}^-\cong v_r {\frac{H}{r}}\left[ \frac{38}{9}aT^4 +
\frac{3}{8}\frac{\rho k T}{m_n}(1+X_{\rm nuc})\right],
 \ee
where the first term includes both the entropy density of
radiation and that of neutrinos.

The neutrino cooling rate integrated over $H$ in the optically
thin limit is just
 \be
Q_\mathrm{\nu}^-=(\dot{q}_{eN}+\dot{q}_{\nu_i\bar{\nu}_i}+\dot{q}_{\rm
brem}+q_{\rm plas})H,
 \ee
  where $\dot{q}_{eN}$ is the cooling rate
due to electron or positron capture by a nucleon,
$\dot{q}_{\nu_i\bar{\nu}_i}$ is that caused by electron-positron
pair annihilation, $\dot{q}_{\rm brem}$ is the cooling term
corresponding to nucleon-nucleon bremsstrahlung , and
$\dot{q}_{\rm plas}$ is the rate due to plasmon decays, (Kohri et
al. 2005; Di Matteo et al. 2002; Kohri \& Mineshige 2002):
 \be
&\dot{q}_{Ne}&=9.2\cdot 10^{33}\rho_{10}T_{11}^6X_{\rm nuc}{\ \rm erg \ cm}^{-3}{\rm s}^{-1}, \label{qen}\\
&\dot{q}_{\nu_e\bar{\nu}_e}&=3.4 \cdot 10^{33} T_{11}^9{\ \rm erg \ cm}^{-3}{\rm s}^{-1},\label{qee}\\
&\dot{q}_{\nu_\mu\bar{\nu}_\mu}&=q_{\nu_\tau\bar{\nu}_\tau}=0.7
\cdot
10^{33} T_{11}^9{\rm \ erg \ cm}^{-3}{\rm s}^{-1},\label{qeex}\\
&\dot{q}_{\rm brem}&=1.5\cdot 10^{27} \rho_{10}^2T_{11}^{5.5} {\rm
erg \ cm}^{-3}{\rm s}^{-1},\label{qbre}\\  &\dot{q}_{\rm
plas}&=3\cdot 10^{32} T_{11}^9e^{-\gamma_{\rm p}}(\gamma_{\rm
p}^6+\gamma_{\rm p}^7+\frac{\gamma_{\rm p}^8}{2}){\rm erg \
cm}^{-3}{\rm s}^{-1}.\label{qpla}
 \ee
In these equations,
 \be\gamma_{\rm p}&=&5.565\cdot
10^{-2}\sqrt{(\pi^2+3\eta_e^2)/3},\\
 \eta_e&=&\mu_e/(kT),
 \ee and
the electron chemical potential $\mu_e$ is obtained by solving Eq.
(26) of Kohri \& Mineshige (2002).

Without assuming neutrino transparency, it is necessary to
consider the inverse processes to (\ref{qen})-(\ref{qpla}) that
produce absorption of neutrinos, as well as the scattering with
nucleons that may prevent the free escape of neutrinos. The
absorptive optical depths for the three neutrino flavors are
(Kohri et al. 2005)
 \be
 &\tau_{a,\nu_e}&=\frac{(\dot{q}_{eN}+\dot{q}_{e^-e^+\rightarrow \nu_e\bar{\nu}_e}+\dot{q}_{\rm
brem}+q_{\rm plas})H}{(7/2)\sigma T^4}\\
&\tau_{a,\nu_\mu}&=\tau_{a,\nu_\tau}=\frac{(\dot{q}_{e^-e^+\rightarrow
\nu_\mu\bar{\nu}_\mu}+\dot{q}_{\rm brem})H}{(7/2)\sigma T^4},
 \ee
whereas for the scattering optical depth, we use the expression
given by Di Matteo et al (2002),
 \be
 \tau_{s,\nu_i}\cong 2.7\cdot 10^{-7}T_{11}^2\rho_{10}H.
 \ee
Then, adopting a simplified model for the neutrino transport
(Popham \& Narayan 1995), the neutrino flux integrated over a half
thickness is
 \be Q_{\nu}^-=\sum_i{\frac{7/8 \sigma
 T^4}{(3/4)(\tau_i/2+1/\sqrt{3}+1/(3\tau_{a,i}))}},
 \ee
where $\tau_i=\tau_{a,i}+\tau_{s,i}$. We employ this last
expression with which the neutrino emission is correctly obtained
in situations with both small and large optical depths.

For the purposes of this paper, we need to work out the structure
of a steady-state accretion disk around a rotating black hole with
a mass $M_{\rm bh}=3M_\odot$, a dimensionless spin parameter
$a_*=0.9$ ($a_*=1$ would be a maximally rotating black hole), a
viscosity parameter $\alpha=0.1$, and an accretion rate between
$\dot{M}=0.1-1M_\odot $s$^{-1}$. It is necessary to introduce
general relativistic correction factors to some of the equations
presented above. These factors are (Riffert \& Herold 1995):
 \be
 &A&=1-\frac{2GM_{\rm bh}}{c^2r}+\left(\frac{GM_{\rm
 bh}a_*}{c^2r}\right)^2\\
 &B&=1-\frac{3GM_{\rm bh}}{c^2r}+2a_*\left(\frac{GM_{\rm bh}a_*}{c^2r}\right)^{3/2}\\
 &C&=1-4a_*\left(\frac{GM_{\rm bh}a_*}{c^2r}\right)^{3/2}+3\left(\frac{GM_{\rm
 bh}a_*}{c^2r}\right)^2\\
 &D&=\int_{r_{\rm ms}}^r{\frac{\frac{x^2c^4}{8G^2}-\frac{3xM_{\rm bh}c^2}{4G}+\sqrt{\frac{a_*^2M_{\rm}^3c^2x}{G}}-\frac{3a_*^2M_{\rm bh}^2}{8} }
 {\frac{\sqrt{r x}}{4}\left(\frac{x^2c^4}{G^2} -\frac{3xM_{\rm
 bh}c^2}{G}+2\sqrt{\frac{a_*^2M_{\rm}^3c^2x}{G}}\right)}
 }dx,
 \ee
Our equation (\ref{masscnsv}) remains valid, while hydrostatic
equilibrium in the vertical direction leads to a corrected
expression for the half thickness of the disk (Riffert \& Herold
1995):
 \be
H \simeq \sqrt{\frac{Pr^3}{\rho G M_{\rm bh}}}\sqrt{\frac{B}{C}}
\label{Hok}.
 \ee
The viscous shear $f_\phi$ of Eq. (\ref{angcnsv}) is also
corrected by
 \be
f_\phi=\alpha P \sqrt{\frac{A}{BC}}=\frac{\dot{M}}{4\pi
H}\sqrt{\frac{GM_{\rm bh}}{r^3}}\frac{D}{A},\label{angcnsvok}
 \ee
and the energy balance is affected through the correction in the
heating rate,
 \be
 \dot{Q}^+_{\rm vis}=\frac{3 \dot{M}G  M_{\rm bh} }{8\pi H
 r^3}\frac{D}{B}.\label{Qhok}
 \ee

With a fixed radius, we numerically solve for $H$ and $P$ Eqs.
(\ref{Hok}) and (\ref{angcnsvok}), which depend only on $\rho$,
and using the equation of state (\ref{Pest}) we get $T(\rho)$.
Then, when the equality (\ref{bal}) is satisfied, the
corresponding values of $T$ and $\rho$ are selected for that
radius. In Fig. 1 we show the temperature, density, and half
thickness profiles for accretion rates of $0.1M_\odot$s$^{-1}$ and
$1M_\odot$s$^{-1}$, and in Fig. 2 we present the neutrino cooling
parameter $f_\nu\equiv Q_\nu^-/Q^+_{\rm vis}$ and the advection
parameter $f_\nu\equiv Q_{\rm adv}^-/Q^+_{\rm vis}$. We observe
that for a wide range of disk radii the most important cooling
mechanism is neutrino emission with advection in the second place,
dominating the latter only in the first $\sim 2$ km from the inner
radius ($R_{\rm ms}\simeq 10.6$ km) as a consequence of the
relativistic corrections made. Elsewhere, up to $\sim 200-750$ km
depending on the accretion rate, neutrino cooling dominates.
\begin{figure}[ht]
\includegraphics[trim = 0mm 5mm 0mm 7mm, clip, width=8cm,angle=0]{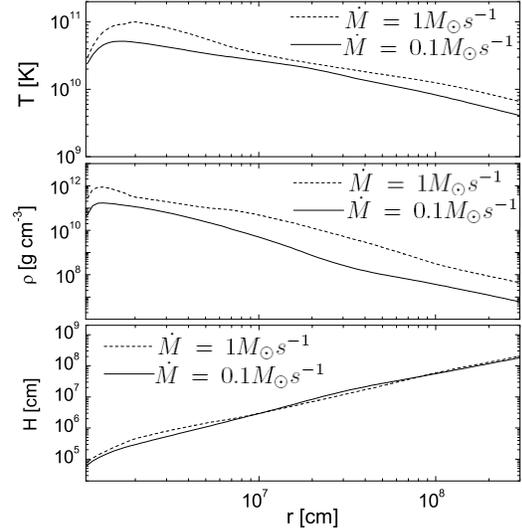}
\caption{Disk temperature profile in the upper panel, density
profile in the middle panel, and half thickness in the lower
panel. Dashed line: $\dot{M}=1M_\odot \rm s^{-1}$, solid line:
$\dot{M}=0.1M_\odot \rm s^{-1}$.}
\end{figure}

\begin{figure}[ht]
\includegraphics[trim = 0mm 6mm 0mm 7mm, clip, width=8cm,angle=0]{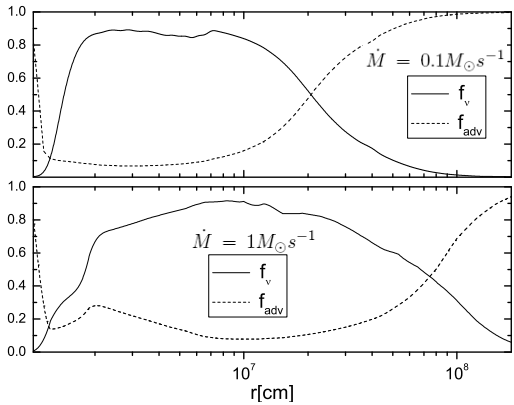}
\caption{Neutrino cooling parameter $f_\nu$ (solid line) and
advection parameter $f_{\rm adv}$ (dashed line) in the upper panel
for $\dot{M}=0.1M_\odot \rm s^{-1}$ and in the lower panel for
$\dot{M}=1M_\odot \rm s^{-1}$.}
\end{figure}


\section{Spin-induced precession of accretion disks}

A Kerr black hole has been found to produce the Lense-Thirring
effect (Lense \& Thirring 1918), or dragging of inertial frames,
which induces the precession of a particle orbit that happens to
be inclined with respect to the black hole equator. The angular
momentum of a Kerr black hole of mass $M_\mathrm{bh}$ is $J=G
M_\mathrm{bh}|a_*|/c$, where $a_*$ is the dimensionless spin
parameter. In such a case, an inclined particle orbit has been
found to precess with a frequency $w_{\rm LT}=GJ/(c^2r^3)$
(Wilkins 1972).

For the candidate central engines of GRB, namely collapsars, the
common envelope evolution of a black hole that causes the tidal
disruption of a companion helium core, or other types of merging
events, the formation of an accretion disk inclined with respect
to the Kerr black hole equator cannot be ruled out. In the last
two cases, it is clear that the black hole angular momentum
$\vec{J}$ does not need to be aligned with the one of its
companion, and in the case of collapsars, when the core collapses
to form a black hole, it could happen that the rest of the rapidly
rotating star may not be symmetrically distributed in the presence
of magnetic fields, leading to the formation of an accretion disk
whose angular momentum $\vec{L}$ would not be exactly aligned with
that of the black hole.

In any of the above situations, the inclined disk is also expected
to undergo the Bardeen-Patterson effect (Bardeen \& Patterson
1975), that results from the action of viscous torques together
with the Lense-Thirring effect. This causes an alignment of the
inner parts of the disk with the black hole equator, which arises
up to a certain transition radius that depends on the midplane
Mach number of the disk (Nelson \& Papaloizou, 2000). In the case
considered here, the midplane Mach number is $\mathcal{M} < 5 $,
which corresponds to essentially no warping of the disk according
to Nelson \& Papaloizou (2000). Also, the disk is expected to
precess as a solid body if the sound crossing timescale (here
$\tau_{\rm cross}<3.2 \cdot 10^{-2}$s) is shorter than the
precession period (Papaloizou \& Terquem 1995). For precession
periods $\tau_{\rm p}>\tau_{\rm cross}$, we will then assume rigid
body precession, and since there is a torque causing the
precession, considering precession alone is actually an
approximation, meaning that the same torque should also cause a
nutation motion to develop
(Goldstein et al. 2002).

Assuming then that the disk supports a strong magnetic field that
threads the central black hole, rotational energy can be extracted
from the black hole due to mechanisms such as those proposed by
Blandford \& Znajek (1977) or by Blandford \& Payne (1982). As the
magnetic field is expected to be anchored in the disk, this would
lead to the formation of jets perpendicular to the midplane of the
disk, so that the precession and nutation of the disk
automatically implies the precession and nutation of jets. Then,
to kinematically describe the jet motion, we consider the angular
evolution of its spherical angles $\theta_\mathrm{jet}(t)$ and
$\phi_\mathrm{jet}(t)$ as in Portegies et al. (1999),
 \be \label{kine}
   \phi_{\mathrm{jet}}(t)&=& \phi_\mathrm{jet}(0)+\Omega_{\mathrm{p}}t +
   \frac{\Omega_{\mathrm{p}}}{\Omega_\mathrm{n}} \sin(\Omega_{\mathrm{n}}t),\nonumber \\
   \theta_{\mathrm{jet}}(t)&=&\theta^0_{\mathrm{jet}}+\frac{\Omega_{\mathrm{p}}}{\Omega_{\mathrm{n}}}\tan
   \theta^0_{\mathrm{jet}}\cos(\Omega_{\mathrm{n}}t),
 \ee
where $\Omega_{\mathrm{p}}=2\pi/\tau_\mathrm{p}$, and
$\Omega_{\mathrm{n}}=2\pi/\tau_\mathrm{n}$ are the precession and
nutation angular frequencies with periods $\tau_\mathrm{p}$ and
$\tau_\mathrm{n}$, respectively. A sketch of the situation is
shown in Fig. 3.

\begin{figure}
\centering
\includegraphics[trim = 10mm 160mm 100mm 13mm, clip, width=8cm]{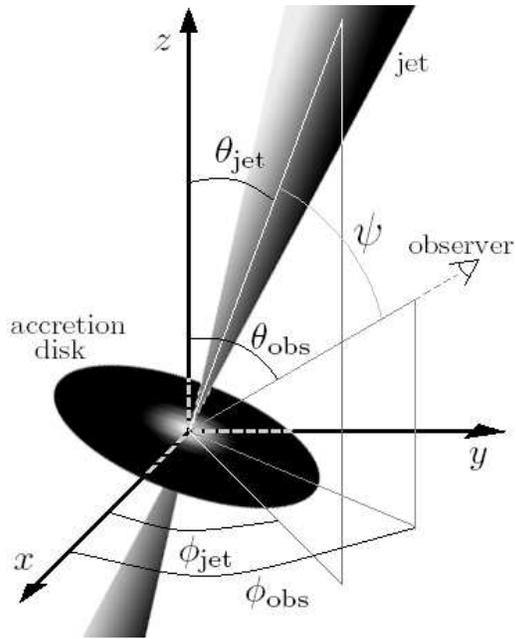}
\caption{Schematic picture of
the precessing disk and jet. The observer is at infinity in the
direction signaled.}
\end{figure}

Considering the above expressions (\ref{kine}), we intend to
relate the possible precession and nutation periods to some
parameter of a given disk, e.g. the surface density $\Sigma(r)$.
We do this in a similar way to Caproni et al. (\cite{caproni}),
but here nutation will also be considered along with precession.
Without nutation, the precession period $\tau_\mathrm{p}$ can be
estimated to be
 \be
  \tau_p=\int_{0}^{2\pi} \frac{L}{T}\sin \theta d\phi = 2
  \pi \sin \theta \frac{L}{T},
 \ee
where the magnitudes of the disk angular momentum $L$ and of the
precessional torque $T$ applied to the disk are
 \be L &=& 2
 \pi\int_{R_\mathrm{ms}}^{R_\mathrm{out}} \Sigma(r) \Omega_\mathrm{k}(r) r^3 \ dr \\
 T&=&4\pi^2 \sin{\theta}
 \int_{R_\mathrm{ms}}^{R_\mathrm{out}} \Sigma(r) \Omega_\mathrm{k}(r) \nu_{p,\theta}(r)
 r^3 \ dr.
 \ee
Here, \be \Omega_\mathrm{k}(r)=\frac{c^3}{GM_\mathrm{bh}} \left[
\left( \frac{r}{R_\mathrm{g}} \right)^{3/2}+a_*  \right]^{-1} \ee
is the relativistic Keplerian angular velocity,
$R_\mathrm{g}=GM_\mathrm{bh}/c^2$ the gravitational radius, and
 \be
 \nu_{p,\theta}=\frac{\Omega_k(R)}{2\pi} \left[1 - \sqrt{1\mp
4\sqrt{\frac{R_\mathrm{g}a_*^2}{r}}+3 \left(\frac{R_{\rm
g}a_*}{r}\right)^2}\right]
 \ee
the nodal frequency that comes from perturbing a circular orbit in
the Kerr metric (Kato 1990). The precessing part of the disk ends
at an outer radius $R_{\mathrm{out}}$, extending from an inner
radius $R_\mathrm{ms}= \xi_{\rm ms}R_\mathrm{g}$, where \be
\xi_\mathrm{ms}=3 +A_2 \mp [(3-A_1)(3+A_1+2A_2)]^{1/2}, \ee with
\be A_1=1+(1-a_*^2)^{1/3}[(1+a_*)^{1/3}+(1-a_*)^{1/3}], \ee and
\be A_2=(3a_*^2+A_1^2)^{1/2}, \ee where the minus sign in
$\xi_{\rm ms}$ corresponds to prograde motion ($a_*>0$), whereas
the plus sign corresponds to retrograde motion ($a_*<0$). Note
that for $a_*=0.9$, $R_{\rm ms}\simeq 1.1\cdot 10^{6}$cm.

With both nutation and precession, according to (\ref{kine}), \be
\phi(\tau_\mathrm{p})=2\pi+(\tau_\mathrm{n}/\tau_\mathrm{p})
\sin(2\pi\tau_\mathrm{p}/\tau_\mathrm{n}), \ee so we estimate \be
\tau_p&=&\int_{0}^{\phi(\tau_\mathrm{p})} \frac{L}{T}\sin \theta(t)
d\phi \\ \tau_p&=& \frac{2\pi G M_\mathrm{bh}}{c^3}\left[ 2\pi
+\frac{\sin(2\pi R_\Omega)}{R_\Omega}\right]\times \nonumber\\ & &
\frac{\int_{\xi_\mathrm{ms}}^{\xi_\mathrm{out}}
\Sigma(\xi)\left[\Gamma(\xi)\right]^{-1}\xi^{3}d\xi}
{\int_{\xi_\mathrm{ms}}^{\xi_\mathrm{out}} \Sigma(\xi)\Psi(\xi)
 \left[\Gamma(\xi)\right]^{-2}\xi^{3}d\xi},  \label{P2M}
\ee
where $R_\Omega=\frac{\tau_\mathrm{p}}{\tau_\mathrm{n}}$,
$\xi=r/R_\mathrm{g}$, $\Gamma(\xi)=\xi^{1/2}+a_*$, and
\be
\Psi(\xi)=1-(1\mp 4a_*\xi^{-3/2} + 3a_*^2\xi^{-2})^{1/2}.
\ee

\section{Results}

We intend to evaluate the possible precession and nutation periods
that are consistent with the typical parameters of an accretion
disk     
which is to power a GRB, i.e. $M_\mathrm{bh}=3M_\odot$,
 $\dot{M}=0.1 - 1 \; M_\odot \mathrm{s}^{-1}$, $\alpha = 0.1$, and $a_*
\sim 0.9$.
Then, taking into account a disk surface density obtained as in
Sect. 2 by $\Sigma(r)=2\rho(r) H(r)$,
 we compute the possible precession periods using (\ref{P2M}) with
$a_*=0.9$ for a wide range in $R_\mathrm{out}$($\sim 10^6-10^9$cm)
and also for the relative fraction
$R_\Omega=\tau_\mathrm{p}/\tau_\mathrm{n}$ varying between $10^{-3}$
and $10$. We thus obtain the precession period $\tau_\mathrm{p}$ as
a function of both the outer radius $R_\mathrm{out}$ of the
accretion disk and the fraction $R_\Omega$. In Fig. 4 we show the
results for an accretion rate of $\dot{M}=0.1 \; M_\odot
\mathrm{s}^{-1}$.

\begin{figure}[h]
\vspace{-0.5cm} \hspace{-0.4cm}
\centering
\includegraphics[trim = 0mm 2mm 6mm 0mm, clip, width=7.8cm,angle=0]{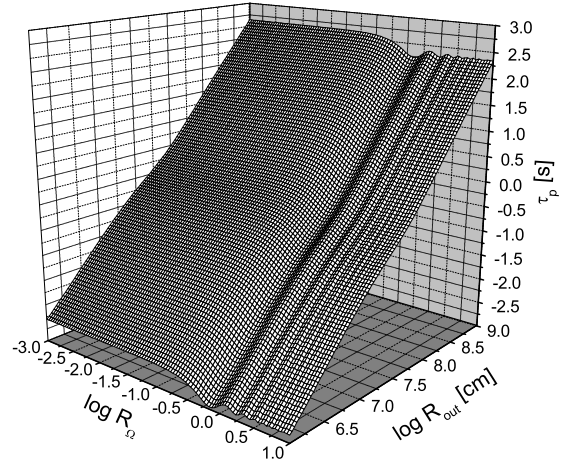}
\caption{Precession period vs $R_\mathrm{out}$ and
$R_\Omega=\tau_\mathrm{n}/\tau_\mathrm{p}$ for $\dot{M}=0.1M_\odot
s^{-1}$.}
\end{figure}

Mac Fayden \& Woosley (1999) note that a steady disk may form
within the collapsar model at $r \sim 2 - 3 \cdot 10^7
\mathrm{cm}$, so looking at $R_\mathrm{out}$ of that order may
give a rough idea of the possible precession periods. Then, by
fixing $R_\Omega$ at $10$, $0.1$ and $0.01$, we find the
precession period $\tau_\mathrm{p}$ as a function of the outer
radius $R_\mathrm{out}$ for nutation periods of
$\tau_\mathrm{n}=\tau_\mathrm{p}/10$, $\tau_\mathrm{p}/0.1$, and
$\tau_\mathrm{p}/0.01$, respectively.
The results are shown in Fig. 5 for $\dot{M}=0.1 M_\odot {\rm
s}^{-1}$ and $\dot{M}=1 M_\odot {\rm s}^{-1}$.
\begin{figure}[h]
\hspace{-0.4cm}
\includegraphics[trim = 0mm 5mm 0mm 5mm, clip, width=9cm,angle=0]{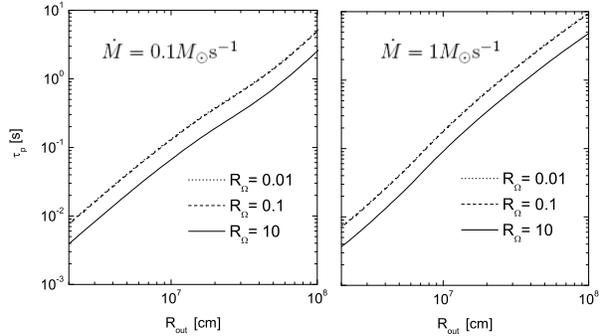}
\caption{Precession period as a function of the disk outer radius
for $R_\Omega=\left\{0.01,0.1,10\right\}$. Left panel shows the
results for $\dot{M}=0.1M_\odot {\rm s}^{-1}$, and the right panel
for $\dot{M}=1M_\odot {\rm s}^{-1}$.}
\end{figure}


It has been found that neutrino emission is important in the inner
regions of the disk where very high temperatures and densities are
achieved (Di Matteo 2002, Khori \& Mineshige 2002), in agreement
with our results. In particular, from Fig. 2, we observe that
neutrino emission dominates until a radius of $\sim 2 \cdot
10^{7}$cm for  $\dot{M}=0.1M_\odot$s$^{-1}$ and up to $7.5 \cdot
10^{7}$cm for $\dot{M}=1M_\odot$s$^{-1}$, so these neutrino-cooled
accretion disks may precess approximately with periods (see Fig.
5)

 \be
   \left. \tau_{\rm p} \right|_{0.1M_\odot {\rm s}^{-1}}  &<& 0.38 \mbox{ s  for } R_\Omega \simeq 0.01, \\
   \left. \tau_{\rm p} \right|_{0.1M_\odot {\rm s}^{-1}}  &<& 0.41 \mbox{ s  for } R_\Omega \simeq 0.1, \\
   \left. \tau_{\rm p} \right|_{0.1M_\odot {\rm s}^{-1}} &<& 0.2 \mbox{ s  for } R_\Omega \simeq
   10,
 \ee
and
 \be
   \left. \tau_{\rm p}\right|_{1M_\odot {\rm s}^{-1}}  &<& 6.21 \mbox{ s  for } R_\Omega \simeq 0.01, \\
   \left. \tau_{\rm p}\right|_{1M_\odot {\rm s}^{-1}}  &<& 6.04 \mbox{ s  for } R_\Omega \simeq 0.1, \\
   \left. \tau_{\rm p}\right|_{1M_\odot {\rm s}^{-1}} &<& 3.12 \mbox{ s  for } R_\Omega \simeq
   10.
 \ee
We note that, for example, if $R_\Omega=0.1$, intermediate
precession periods of $0.41$ s and $6.04$ s can also be achieved
with intermediate values of $\dot{M}$ between $0.1M_\odot$s$^{-1}$
and
 $1M_\odot$s$^{-1}$ for different outer radii between $2\cdot10^{7}$ cm and  $7.5\cdot10^{7}$ cm.

\begin{figure}[h]
\hspace{-1cm}
\centering
\includegraphics[trim = 0mm 0mm 3mm 25mm, clip, width=6.5cm,angle=-90]{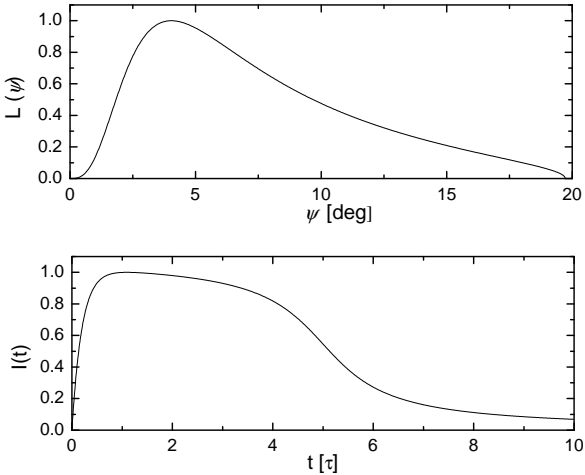}
\caption{Upper panel: Jet luminosity $L$ as a function of the
angle $\psi$ between the jet and the observer. Lower panel:
Intrinsic time dependence of the burst.}
\end{figure}

\begin{figure}[h]
\centering
\includegraphics[trim = 0mm 5mm 0mm 5mm, clip, width=8.5cm,angle=0]{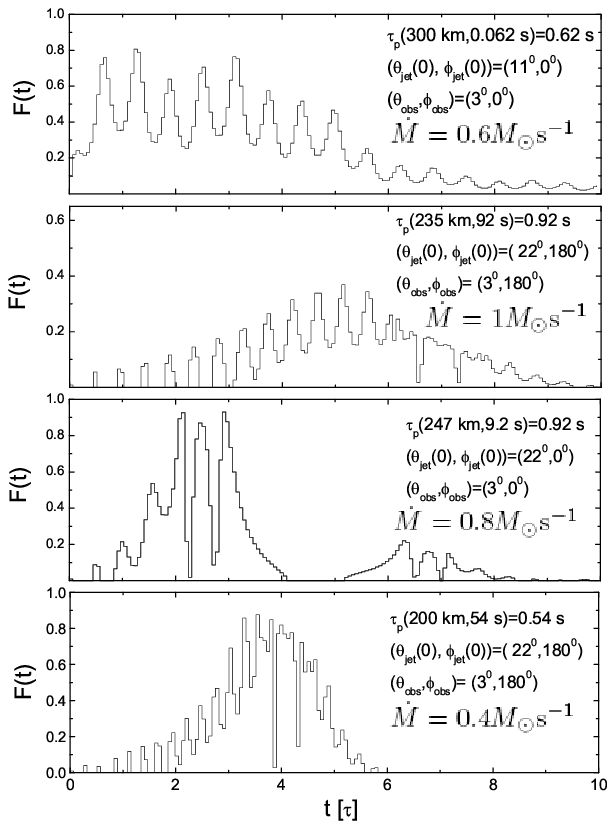}
\caption{Some possible GRB light curves for precessing
neutrino-cooled disks with $0.6 M_\odot {\rm s}^{-1}$(upper
panel), $1 M_\odot {\rm s}^{-1}$(second panel), $0.8 M_\odot {\rm
s}^{-1}$(third panel) and $0.4 M_\odot {\rm s}^{-1}$(bottom
panel). The precession period is indicated as $\tau_{\rm p}(R_{\rm
out},\tau_{\rm n})$, the jet angles at $t=0$ as $(\theta_{\rm
jet}(0),\phi_{\rm jet}(0))$ and the observer angles as
$(\theta_{\rm obs},\phi_{\rm obs})$.}
\end{figure}
\begin{figure}[ht]
\includegraphics[trim = 0mm 5mm 0mm 5mm, clip, width=8.5cm,angle=0]{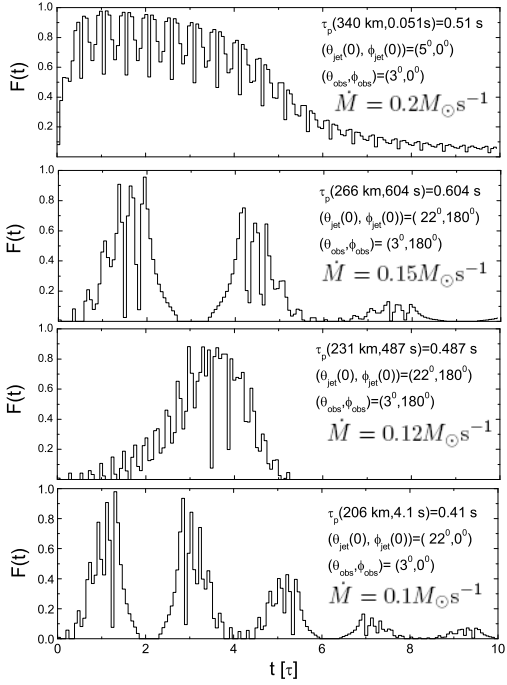}
\caption{Some possible GRB light curves for precessing
neutrino-cooled disks with $0.2 M_\odot {\rm s}^{-1}$(upper
panel), $0.15 M_\odot {\rm s}^{-1}$(second panel), $0.12 M_\odot
{\rm s}^{-1}$(third panel) and $0.1 M_\odot {\rm s}^{-1}$(bottom
panel). The precession period is indicated as $\tau_{\rm p}(R_{\rm
out},\tau_{\rm n})$, the jet angles at $t=0$ as $(\theta_{\rm
jet}(0),\phi_{\rm jet}(0))$ and the observer angles as
$(\theta_{\rm obs},\phi_{\rm obs})$.}
\end{figure}

 Now, to work out possible GRB light curves consistent with such
precessing neutrino-cooled disks, we proceed as do Portegies et
al. (1999) and assume the production of jets perpendicular to the
plane of the disk. The spin angular momentum of the black hole
$\vec{J}$ is along the $z$-axis, we fix the observer at the angles
$(\theta_\mathrm{obs},\phi_\mathrm{obs})$, and then let the jet
angles $(\theta_\mathrm{jet}(t),\phi_\mathrm{jet}(t))$ evolve
according to (\ref{kine}). The angle between the observer and the
jet is $\psi=\cos^{-1}(\hat{r}_\mathrm{obs}\cdot
\hat{r}_\mathrm{jet})$ (see Fig. 3), and, following Portegies et
al. (1999), we implement the distribution of the corresponding
Poynting flux $L(\psi)$ normalized to unity at its maximum (Fig.
6, upper panel). Also, we assume an intrinsic time dependence for
the emission $I(t)$, normalized to unity at its maximum as well,
which is inspired by what is expected from a typical explosive
injection, and consists of a fast rise, a plateau, and an
exponential decay (Fig. 6, lower panel). The intrinsic duration
there was considered to be $10$ s, but it could change from burst
to burst. Defining $F(t)\equiv I(t)L(\psi)$, we show in Fig. 7
some possible light curves varying the initial jet angles and the
observer angles, for different values of
$\tau_\mathrm{p}(R_\mathrm{out},\tau_\mathrm{n})$ and for
relatively high accretion rates. More signals are plotted in Fig.
8 for
lower accretion rates. All these curves have been averaged every
$64$ ms to simulate the resolution of the detector.


\section{Discussion}

Inspecting the light curves shown in both Figs. 7 and 8, and many
others that can be generated in a similar way, we see that some
peculiar time profiles can be produced. For instance, a fast-rise
and exponential-decay (FRED) type of burst, with superposed
microstructure with timescales of $\delta t\sim0.1$ s, as it has
been observed in several cases, can be seen in Fig. 8, upper
panel. One interesting type of burst that can not be accommodated
in the usual internal shock models is formed for those so-called
peculiar asymmetric bursts (PABs). These bursts represent around
$\sim 4$ \% of the total sample (Romero et al. 1999) and present
slower risings than decays. In Fig. 7, bottom panel, and Fig. 8,
3rd panel, we show examples of bursts with such a time profile
that can result from precessing jets with long nutation periods.
In such systems the jet crosses the line of sight before
disappearing quite quickly in the observer's frame. For the right
combination of viewing angle, precession, and nutation, this
results into a typical PAB.

Distinguishing the effects of precession from those of multiple
shocks could be more difficult in more conventional light curves
with many peaks. However, an analysis of the spectral variability
might help at this point, since changes due to precession are
essentially geometric and, hence, achromatic. Inverting the
problem, clear identification of precession in a given time
profile of a GRB can be used to infer some characteristics of the
accretion disk, such as its outer radius.

\section{Summary}

We have studied the precession of accretion disks with neutrino
losses, taking into account our results obtained for the disk
structure by adequately correcting the governing equations of
mass conservation, energy balance, hydrostatic equilibrium in the
vertical direction, and angular momentum conservation in the
presence of a Kerr black hole.
Assuming that precession will not significantly alter
the obtained disk structure and that it will continue to be
determined mainly by the neutrino cooling processes, our results
imply that precession and nutation of such disks is possible in
the context of GRB engines, giving rise to a temporal
microstructure that is similar to what is observed in some cases.
In particular, a peculiar behavior that has been observed
consisting of a slow rise and a fast decay (Romero el al. 1999)
can be obtained in situations with long nutation periods where the
jet crosses the line of sight quite suddenly and never reaches it
again.





\begin{acknowledgements}
This work was supported by the Argentinian agencies CONICET (PIP
5375/05 and PIP 477/98), and ANPCyT (PICT 03-13291 and PICT02
03-11311).
\end{acknowledgements}

{}


\begin{thebibliography}{}
\bibitem{}Artemova, I. V., Bj\"{o}rnsson, G. \& Novikov, I. D. 1996, ApJ, 461,
565
\bibitem{}Bardeen, J. M. \& Patterson, J. A. 1975, ApJ, 195, L65
\bibitem{}Belovorodov, A. M., 1998, MNRAS, 297, 739
\bibitem{}Blackman, E. G., Yi, I. \& Field, G. B. 1996, ApJ, 473,
L79
\bibitem{}Blandford, R. D. \& Payne, D. G. 1982, MNRAS, 199, 883
\bibitem{}Blandford, R. D. \& Znajek, R. L. 1977, MNRAS, 179, 433
\bibitem[2004]{caproni}Caproni, A., Mosquera Cuesta, H. J., \& Abraham, Z. 2004, ApJ, 616,
L99
\bibitem{}Daigne, F, \& Mochkovitch, R. 1998, MNRAS, 296, 275
\bibitem{}Di Matteo, T., Perna, R., \& Narayan, R. 2002, ApJ, 579,
706
\bibitem{}Eichler, D., Livio, M., Piran, T.  \& Schramm, D. N. 1989, Nature, 340,
126
\bibitem{}Fargion, D. 1999, in: D. Kieda, M. Salamon, \& B. Dingus (eds.), 26th ICRC, OG2.3.14, p.32 (astro-ph/9906432)
\bibitem{}Fargion, D. 2003, astro-ph/0307348
\bibitem{}Fenimore, F. E., Madras, C. D., \& Nayakshin, S. 1996, ApJ 473, 998
\bibitem{}Fryer, C. L., \& Woosley, S. E. 1998, ApJ, 502, L9
\bibitem{}Goldstein, H., Poole, C. P. \& Safko, J. L. 2002, Classical Mechanics (Addison-Wesley, Massachusetts), 218
\bibitem{}Guetta, D., Spada, M., \& Waxman, E. 2001, ApJ, 557, 399
\bibitem{}Heinz, S. \& Begelman, M. C. 1999, ApJ, 527, L35
\bibitem{}Kato, S. 1990, PASJ 42, 99
\bibitem{}Kobayashi, S., Piran T., \& Sari, R. 1997, ApJ, 490, 92
\bibitem{}Kohri, K. \& Mineshige, S. 2002, ApJ, 577, 311
\bibitem{}Kohri, K., Narayan, R. \& Piran, T. 2005, ApJ, 629, 341
\bibitem{}Lense, J. \& Thirring, H. 1918, Phys. Z., 19, 156
\bibitem{}MacFayden, A. I. \& Woosley, S. E. 1999, ApJ, 524, 262
\bibitem{}Narayan, R., Paczynski, B. \& Piran, T., 1992, ApJ, 395,
L83
\bibitem{}Narayan, R., \& Yi, I. 1994, ApJ, 428,
L13
\bibitem{}Nelson, R. P. \& Papaloizou, J. C. B. 2000, MNRAS, 315, 570
\bibitem{}Novikov, I. D. \& Thorne, K. S. 1973, in Black Holes, Les
Astres Occlus, ed. B. \& C. DeWitt (New York: Gordon \& Breach),
343(NT73)
\bibitem{}Paczynski, B., 1986, ApJ, 308, L43
\bibitem{}Papaloizou, J. C. B. \& Terquem, C 1995, MNRAS, 274, 987
\bibitem{}Popham, R., Woosley, S. E.  \& Fryer, C. 1999, ApJ, 518, 356
\bibitem{}Popham, R. \& Narayan, R. 1995, ApJ, 442, 337
\bibitem{}Portegies-Zwart, S. F., Lee, C. H., \& Lee, H. K. 1999, ApJ, 520,
666
\bibitem{}Riffert, H. \& Herold, H., 1995, ApJ, 450,
508
\bibitem{}Romero, G. E., Torres, D. F., Andruchow, I. \& Anchordoqui, L. A. 1999, MNRAS, 308,
799
\bibitem{}Setiawan, S., Ruffert, M. \& Janka, H-Th., 2004, MNRAS, 352,
753
\bibitem{}Shakura, N. I. \& Sunyaev, R. A. 1973, A\&A, 24, 337
\bibitem{}Shapiro, P. \& Teukolsky, S. 1983, Black Holes, White
Dwarfs, and Neutron Stars, Wiley, New York
\bibitem{}Wilkins, D. C. 1972, Phys. Rev. D, 5, 814
\bibitem{}Woosley, S. E. 1993, ApJ, 405, 273




\end{thebibliography}
\end{document}